\documentclass[10pt,twocolumn,twoside,letterpaper]{IEEEtran}

\usepackage{geometry}
\makeatletter
\def\ps@IEEEtitlepagestyle{
  \def\@oddfoot{\mycopyrightnotice}
  \def\@evenfoot{}
}
\def\mycopyrightnotice{
  {\footnotesize
  \begin{minipage}{\textwidth}
  \centering
  978-1-6654-2113-3/21/\$31.00 \copyright2021 IEEE
  \end{minipage}
  }
}

\usepackage{siunitx}
\usepackage{url}
\usepackage{graphicx}

\begin{document}
\title{Purification Efficiency and Radon Emanation\\of Gas Purifiers used with Pure and Binary Gas Mixtures for Gaseous Dark Matter Detectors
}
%
%
%
\author{K.~Altenm\"{u}ller$^{1}$, J.~F.~Castel$^{1}$, S.~Cebri\'{a}n$^{1}$, T.~Dafn\'{i}$^{1}$, D.~D\'{i}ez-Ib\'{a}\~{n}ez$^{1}$, J.~Gal\'{a}n$^{1}$, J.~Galindo$^{1}$, J.~A.~Garc\'{i}a$^{1}$, I.~G.~Irastorza$^{1}$, I.~Katsioulas$^{2}$, P.~Knights$^{2,*}$, G.~Luz\'{o}n$^{1}$, I.~Manthos$^{2}$, C.~Margalejo$^{1}$, J.~Matthews$^{2}$, K.~Mavrokoridis$^{3}$, H.~Mirallas$^{1}$, T.~Neep$^{2}$, K.~Nikolopoulos$^{2}$, L.~Obis$^{1}$, A.~Ortiz~de~Sol\'{o}rzano$^{1}$, O.~P\'{e}rez$^{1}$, B.~Philippou$^{3}$, R.~Ward$^{2}$
\thanks{$^{*}$p.r.knights@bham.ac.uk}%
\thanks{$^{1}$Centro de Astropart\'{i}culas y F\'{i}sica de Altas Energ\'{i}as (CAPA), Universidad de Zaragoza, 50009 Zaragoza, Spain}%
\thanks{$^{2}$School of Physics and Astronomy, University of Birmingham, B15 2TT, UK}%
\thanks{$^{3}$Department of Physics, University of Liverpool, L69 7ZE, UK}%
\thanks{
Manuscript received December 2, 2021. This project has received funding from the European Union's Horizon 2020 research and innovation programme under the Marie Sk\l{}odowska-Curie grant agreements no 841261 (DarkSphere) and no 845168 (neutronSphere). }}

\maketitle

\pagenumbering{gobble}

\begin{abstract}
Rare event searches require extreme radiopurity in all detector components. This includes the active medium, which in the case of gaseous detectors, is the operating gas. The gases used typically include noble gas mixtures with molecular quenchers. Purification of these gases is required to achieve the desired detector performance, however, purifiers are known to emanate $^{222}$Rn, which is a potential source of background. Several purifiers are studied for their O$_2$ and H$_2$O purification efficiency and Rn emanation rates, aiming to identify the lowest-Rn options. Furthermore, the absorption of quenchers by the purifiers is assessed when used in a recirculating closed-loop gas system. 
\end{abstract}


\section{Introduction}
\label{sec:introduction}
\IEEEPARstart{G}{aseous} detectors provide a method for expanding direct Dark Matter (DM) searches to masses below $2\;\si{\giga\eV}$. Furthermore, Time Projection Chambers (TPCs) filled with gas is a very promising technology to perform directional DM searches. A recent review can be found in Ref.~\cite{Billard:2021uyg}.  
Noble gases, e.g. He, Ne, Ar, and Xe, are often used in gaseous detectors, usually in mixtures with molecular gases including CH$_4$, C$_3$H$_8$, C$_4$H$_{10}$, CO$_2$, N$_2$, which act as quenchers that stabilise proportional counter operation.

Direct DM searches have stringent requirements on the radioactivity of detector components. 
An important source of background is $^{222}$Rn originating from the $^{238}$U decay chain.
Various techniques have been developed to suppress the $^{222}$Rn deposited on materials, and careful material selection allows for lower $^{238}$U chain daughter isotopes contamination. 
$^{222}$Rn is known to be emanated by gas purifiers~\cite{Ogawa:2019ccp, Knights:2019tmx}, which are necessary to reduce gaseous impurities such as O$_{2}$ and H$_{2}$O, which hinder detector performance.

In this work, the $^{222}$Rn emanation rate was studied for several commercial purifiers: the Entegris Getters, Agilent GasClean, and Messer Oxysorb. Custom made purifiers based on CU-0226 (Q-5) for H$_2$O adsorption, 3{\AA} and 4{\AA} molecular sieves were also studied. 
Key properties of interest are the Rn emanation rate, the purification efficiency and the potential for preferential removal of molecular quencher from the gas mixture. 

\section{Experimental Set-Ups}
To test the $^{222}$Rn emanation and purification efficiency of the purifiers, measurements were conducted with two set-ups: a $30\;\si{\centi\meter}$ in diameter spherical proportional counter \cite{Giomataris:2008ap,Katsioulas:2018pyh, Giomataris:2020rna} and the TREX-DM experimental set-up~\cite{Aznar:2017enu, Castel:2019ngt}.
The spherical proportional counter is a large volume $\mathcal{O}(10\;\si{\litre})$ gaseous detector. It is an efficient $^{222}$Rn detector as it can measured $\mathcal{O}(\si{\micro\becquerel})$ rates and monitor its decay~\cite{Savvidis:2010zz}. The spherical proportional counter used for these measurements is installed in the University of Birmingham and uses a single-pass gas purification system. The TREX-DM experiment uses a low-background MicroMegas-based~\cite{Giomataris:1995fq} TPC installed in the Canfranc underground laboratory. TREX-DM uses a gas recirculation system with a Binary Gas Analyser to continuously purify and monitor the gas quality.

\section{$^{222}$Rn emanation from purifiers}
%
%
Measurements have been performed to assess the $^{222}$Rn emanation of a custom-made filter, UOB-F1, compared to a commercial Entegris MC700 902-F Getter 
by measuring the amount of $^{222}$Rn (and progeny) decays observed following a single-filling of the detector. UOB-F1 contained both Q-5 and the molecular sieves in order to remove water and oxygen contamination.  
For each measurement, the $30\;\si{\centi\meter}$ diameter spherical proportional counter was first evacuated to a pressure of maximum $\num{E-6}\;\si{\milli\bar}$. It was then filled to a pressure of $1\;\si{bar}$ with N$_{2}$ gas that had either been passed through one of the purifiers or directly from the bottle, which was $99.999\%$ pure. N$_{2}$ was used as a reference gas because the detector can be operated without a quencher. 

\begin{figure}[!ht]
\centering
\includegraphics[width=0.9\linewidth]{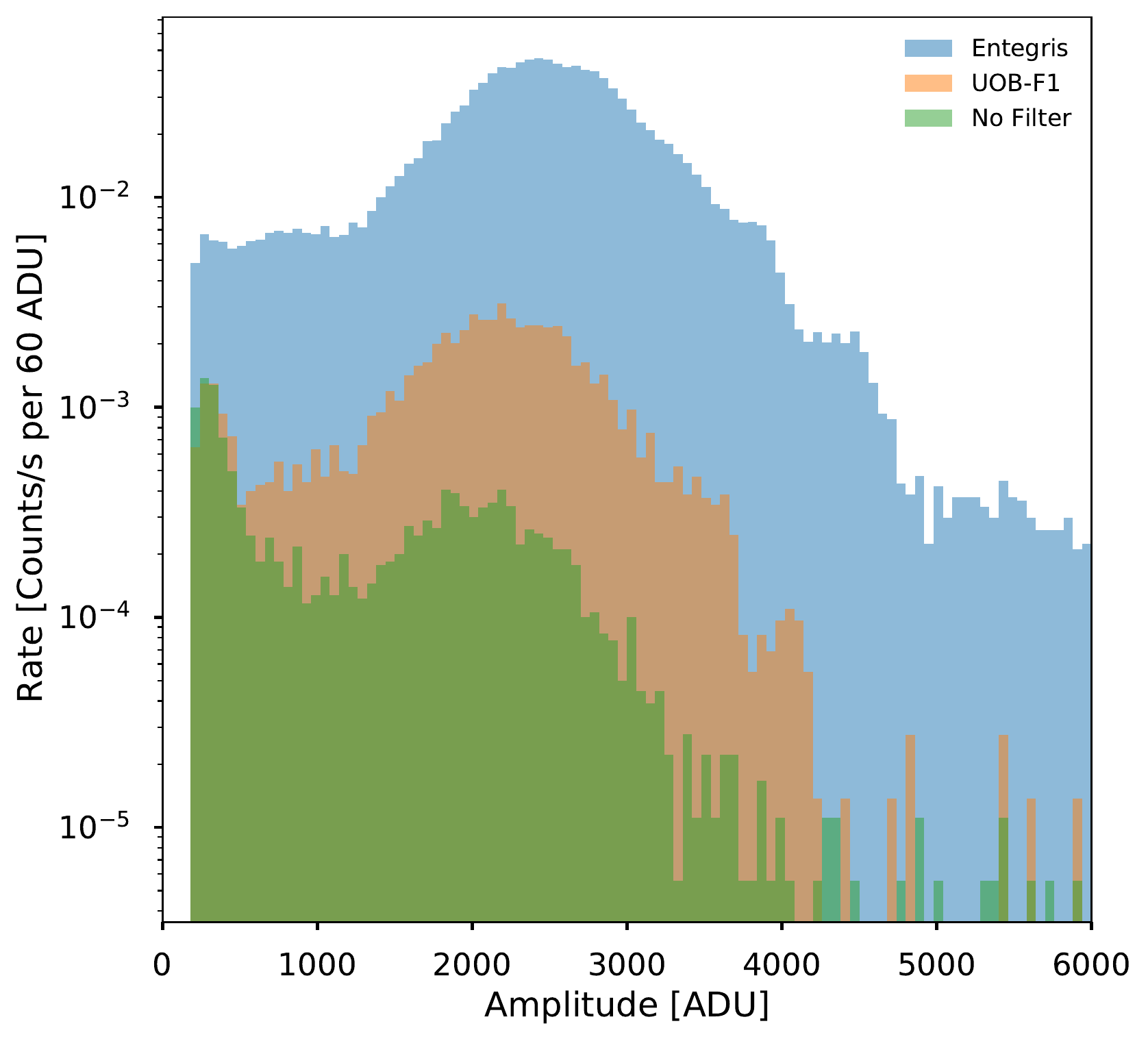}
\caption{
Pulse amplitude measured for $\mathrm{\alpha}$-particles in a spherical proportional counter filled with 1 bar N$_{2}$ gas passed either though an Entegris MC700 902-F Getter, the UOB-F1 purifier or directly from the bottle.
\label{fig:rnComparison}
}
\end{figure}

The measured energy spectrum is shown in Figure~\ref{fig:rnComparison}, with the 
Entegris purifier exhibiting the greatest $\mathrm{\alpha}$-particle rate. 
The $\mathrm{\alpha}$-particle rate with the Entegris purifier was measured to be $1.2\;\si{\Hz}$, whereas with UOB-F1 the rate was $0.06\;\si{ \Hz}$. 
The background $\mathrm{\alpha}$-particle rate was assessed when no purifier was used and was measured to be $0.01\;\si{\Hz}$. 
Thus, UOB-F1 demonstrates a factor 20 reduction in the $^{222}$Rn-emanation rate compared to the Entegris. The Aglient purifier will also be tested using the same method.

\begin{figure}[!h]  
	                    \centering
	      \includegraphics[width=0.45\textwidth]{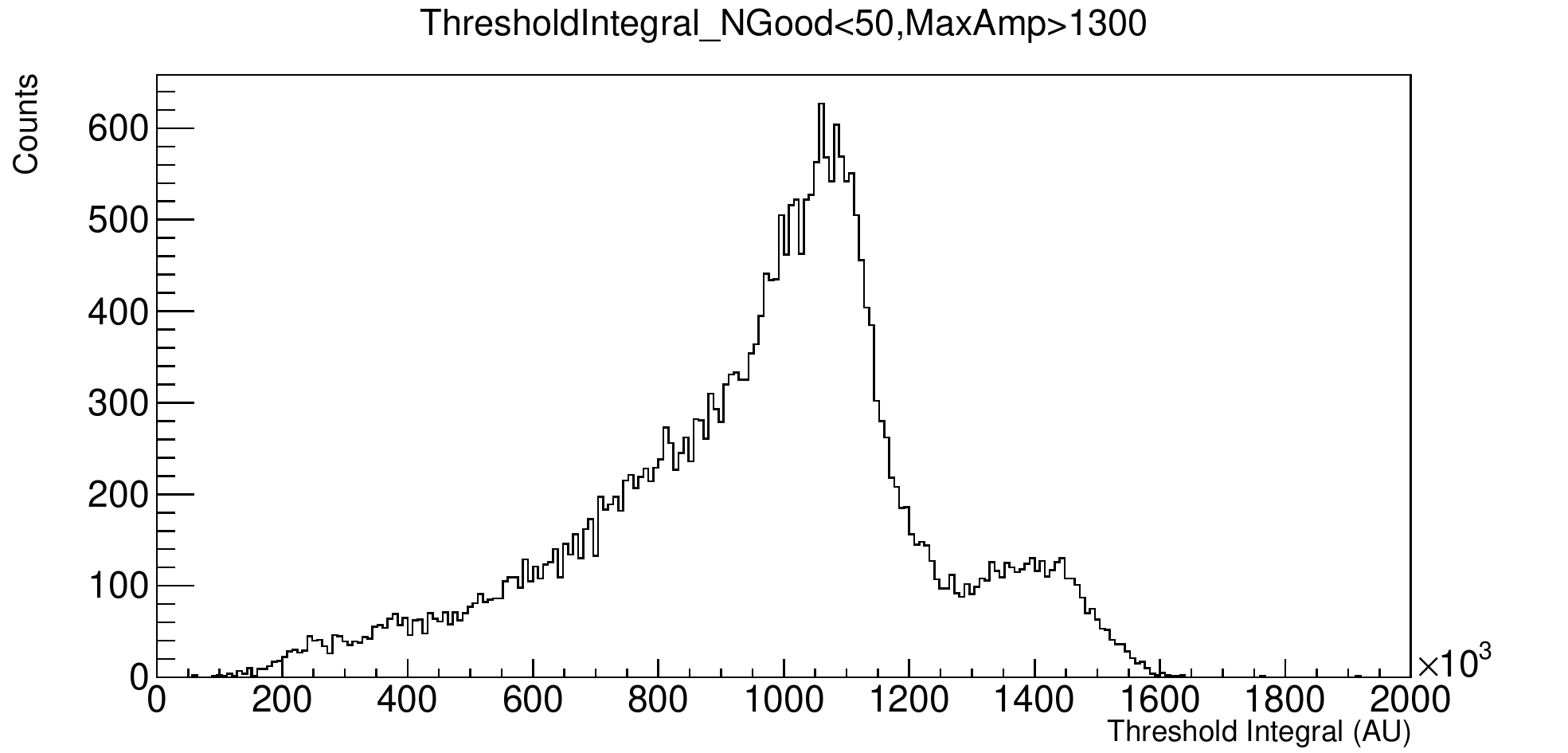}
	      \includegraphics[width=0.45\textwidth]{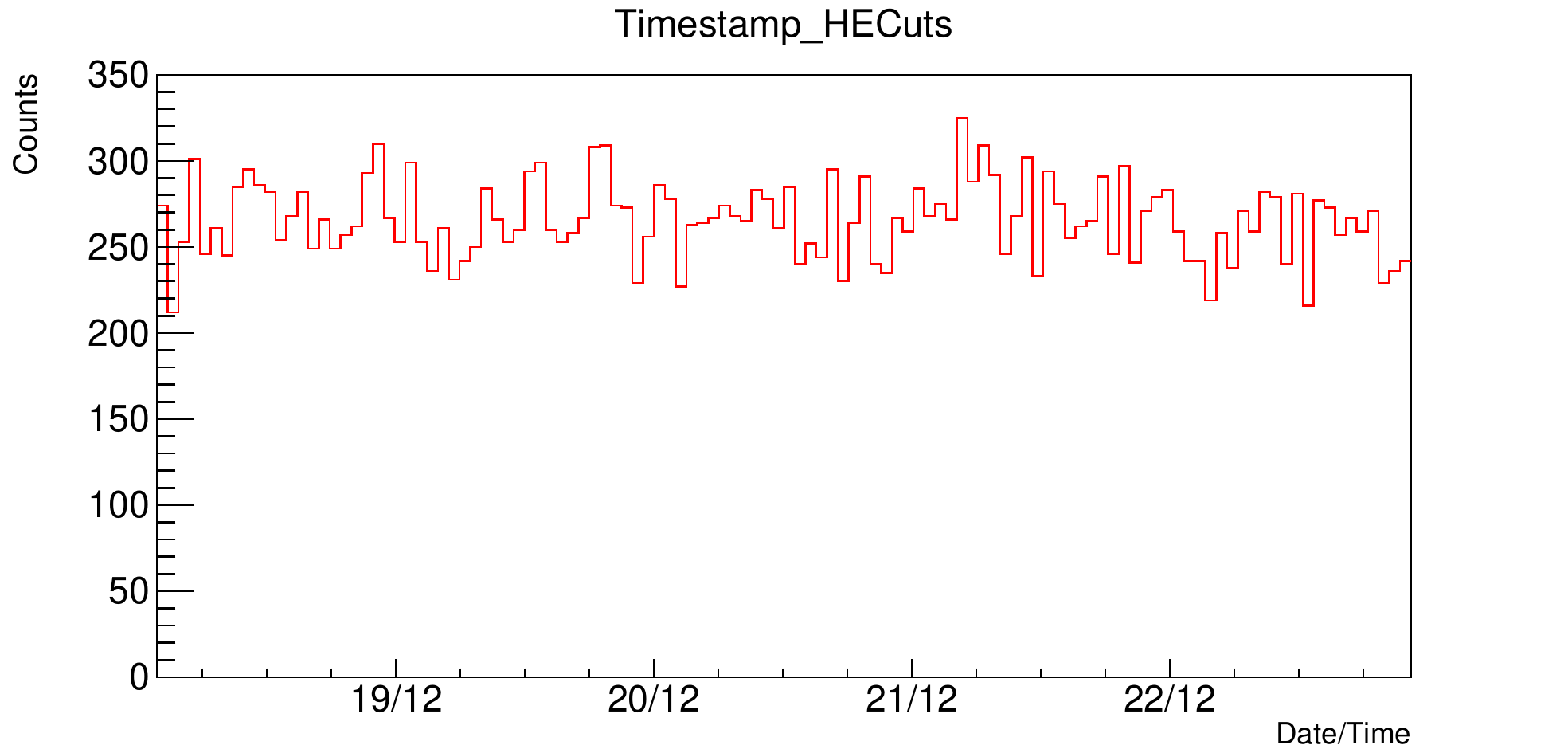}
	\caption{Energy spectrum (upper panel) and number of counts (lower panel) measured by the TREX-DM detector when gas was recirculated through the Agilent GasClean purifier. \label{fig:trexRate}}	      
	    \end{figure}

The $^{222}$Rn emanation of the purifiers was also measured using a recirculation system in the TREX-DM set-up. A Ne:i-C$_{4}$H$_{10}$ ($98\%:2\%$) gas mixture was circulated through the detector and purifier under test, which was either the Entegris, the Aglient GasClean or UOB-F1. The $\mathrm{\alpha}$-particle rate in the detector was monitored for several days until the rate had stabilised, with the amplitude spectrum and number of counts over time shown in Figure~\ref{fig:trexRate} for the Agilent purifier. It was found that over the course of seven days the $\mathrm{\alpha}$-particle rate from the Entegris did not stabilise, while for the Agilent and UOB-F1 it stabilised at $300\;\si{counts\per hour}$ and $150\;\si{counts\per hour}$, respectively. 

To further study the origin of the $^{222}$Rn emanation from UOB-F1, two new purifiers, UOB-F2 and UOB-F3, have been produced which include only the oxygen-removing and water-removing components of UOB-F1, respectively. If one component is identified as emanating more $^{222}$Rn, then the composition of the combined purifier can be adjusted to reduce the emanation rate. 

\section{Purification efficiency}
\begin{figure}[!h]  
	                    \centering
	      \includegraphics[width=0.9\linewidth]{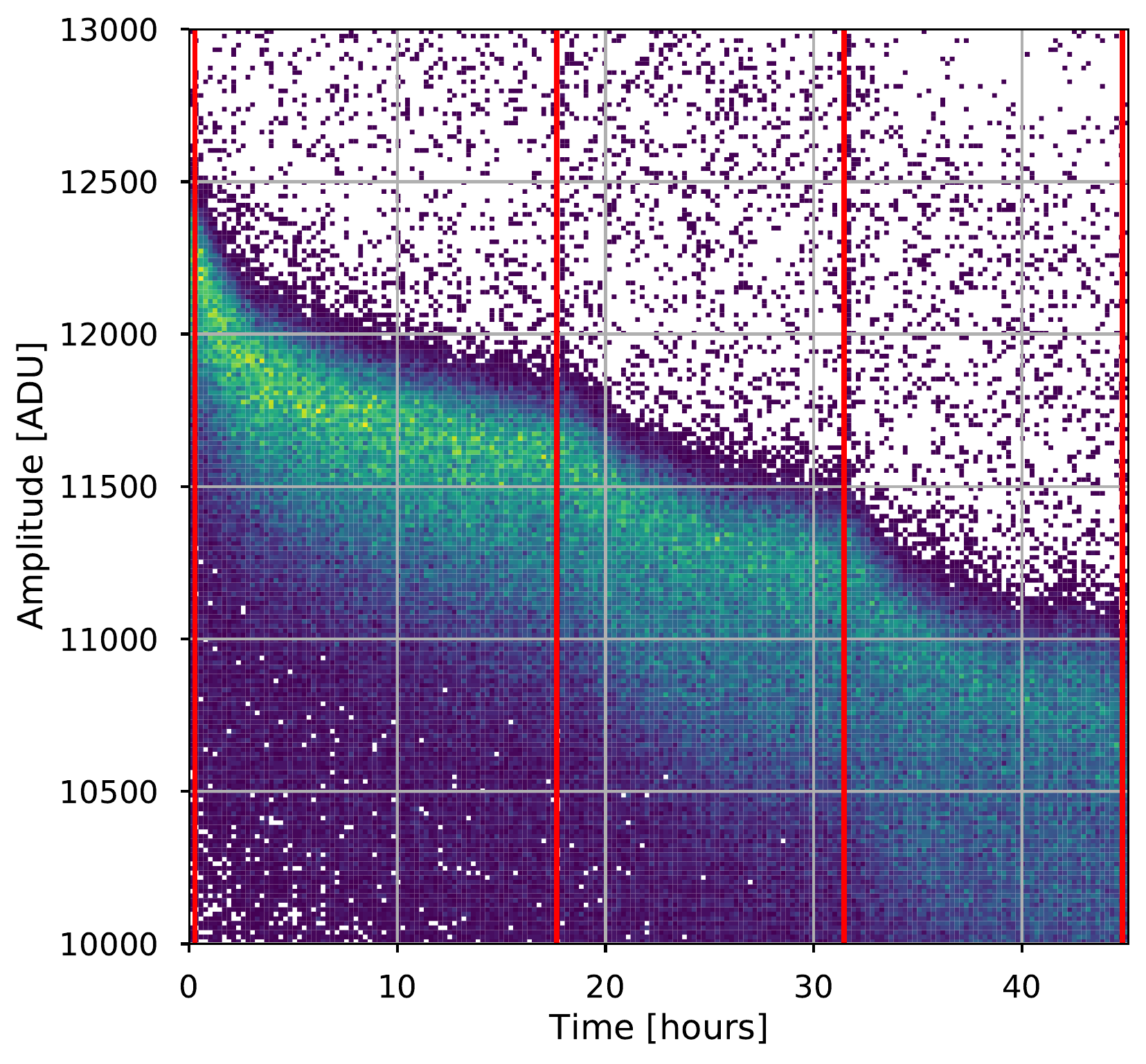}
	      \caption{Measured amplitude versus time for a spherical proportional counter filled with $1\;\si{\bar}$ Ar:CH$_4$ ($98\%:2\%$) passed through UOB-F2. $67.5\;\si{ppm}$ was injected at the times following the first three red lines shown in the figure. \label{fig:airTest}}	      
	      	    \end{figure}

\begin{figure}[!h]  
		     \centering
	      \includegraphics[width=0.9\linewidth]{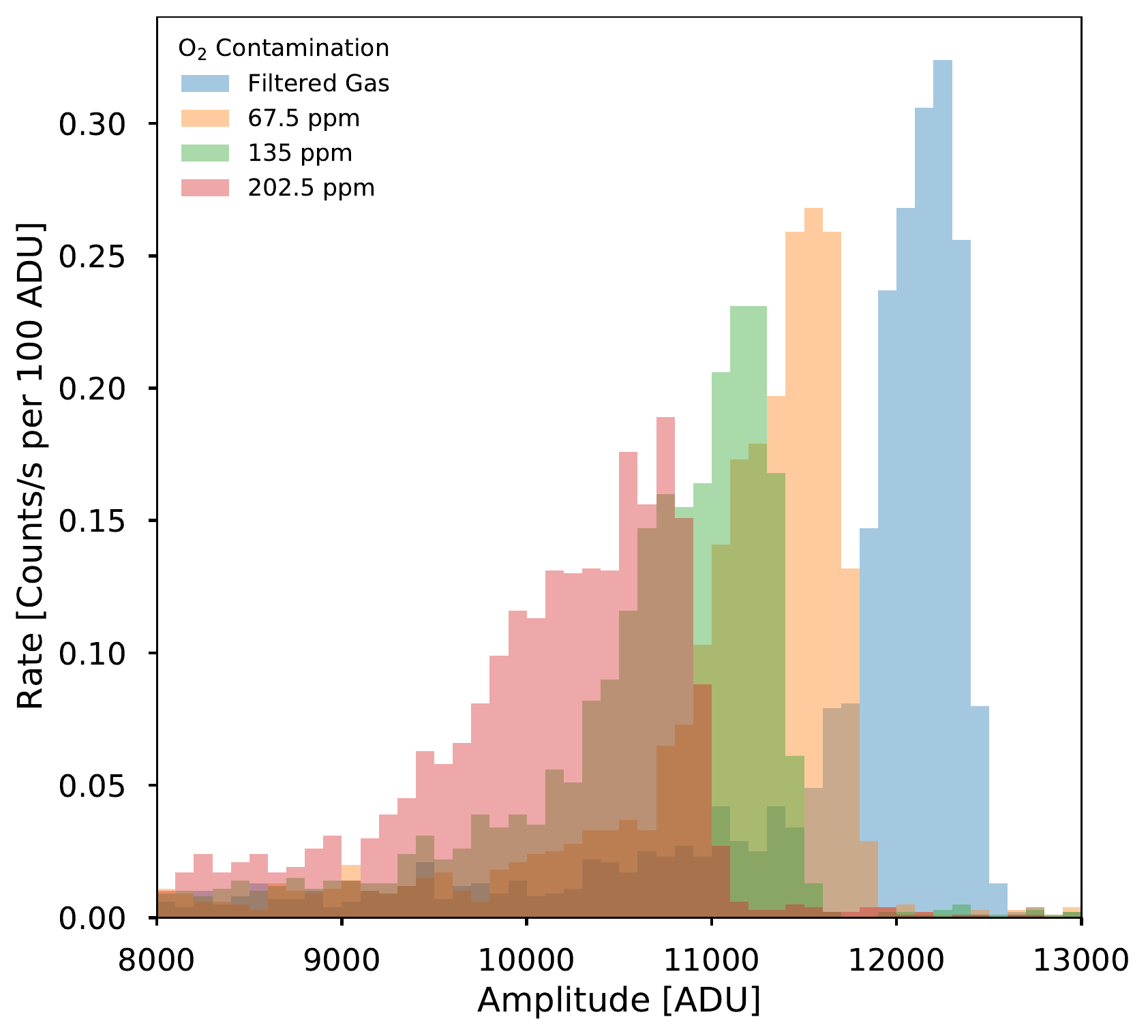}
	\caption{Measured energy spectra for $1\;\si{\bar}$ Ar:CH$_4$ ($98\%:2\%$) passed through UOB-F2 and for three levels of air contamination, quantified in terms of the oxygen contamination. Each histogram corresponds to a red line in Figure~\ref{fig:airTest}. \label{fig:airTest2}}	      
	    \end{figure}

Another key property of the purifiers is their efficiency to remove O$_2$ and H$_2$O from the gas. A change in detector gain and a degradation of the energy resolution are expected for increased amounts of O$_2$ and H$_2$O in the gas. 
The Entegris purifier has previously been used and qualitatively found to achieve satisfactory levels of purification.
Preliminary measurements were conducted with UOB-F2 used to purify $1\;\si{\bar}$ Ar:CH$_4$ ($98\%:2\%$) passed into the spherical proportional counter, and the measured energy spectrum for $5.3\;\si{\mega\eV}$ $\mathrm{\alpha}$-particles from a $^{210}$Po source were used to assess the energy resolution. Figure~\ref{fig:airTest} shows the measured amplitude of the  $5.3\;\si{\mega\eV}$ $\mathrm{\alpha}$-particle as a function of time. After an initial measurement period, $67.5\;\si{ppm}$ of air were injected into the detector, at the time indicated by the first red line at approximately $0.5\;\si{hour}$. Time was then allowed for the air to mix with the gas and the same amount of air was injected again following the times indicated by the second and third  red lines at approximately $17\;\si{hour}$ and $31.5\;\si{hour}$, respectively. The energy spectrum at the times indicated by the red lines in Figure~\ref{fig:airTest}, is shown in Figure~\ref{fig:airTest2}, and shows the degradation of the energy resolution. These results will be combined with the dedicated simulation framework for spherical proportional counters developed at the University of Birmingham~\cite{Katsioulas:2019sui} to infer the amount of oxygen and water present in the purified gas.
%
%
%
\section{Binary Gases}
Gaseous detectors operating with noble gases in proportional mode usually include with a molecular quencher gas. A purifier used with such a gas must not preferentially remove one gas component, as this would change the detectors operation. To study this aspect of each purifier, gas was recirculated through the purifier under test using the TREX-DM set-up. A binary gas analyser in series with the recirculation loop was used to monitor the gas composition over several weeks. This test has been completed with UOB-F1, where a Ne:i-C$_{4}$H$_{10}$ ($98\%:2\%$) gas mixture recirculated for several weeks showed no change to the gas composition. 

\section{Summary}
A survey of the properties of several commercial and custom-made gas purifiers is ongoing to identify a candidate for use with gaseous rare-event search experiments. Results so far indicate that the custom-made UOB-F1 purifier shows promising $^{222}$Rn emanation and qualitatively sufficient oxygen and water purification. It has also been shown to not remove molecular quench gases in binary gas mixtures. 
Measurements are ongoing to complete the tests for all of the purifiers considered in the survey. This includes the $^{222}$Rn emanation in both the single gas filling and recirculation system of each purifier, in particular the UOB-F2 and UOB-F3. The results of this measurement can then be used to further improve the design of UOB-F1. The results of measurements with known amounts of air injected into the purified gas will be used to infer the amount of oxygen and water in the purified gas by comparing with simulations.  
%
%
%

%
%
%
%



\bibliographystyle{IEEEtran}
 \bibliography{mybib} 
%
%
%
%



\end{document}